# On the origins and the historical roots of the Higgs boson research from a bibliometric perspective

A. Barth [1], W. Marx [2], L. Bornmann [3], R. Mutz [4]

[1] FIZ Karlsruhe, Hermann-von-Helmholtz-Platz 1, D-76344 Eggenstein-Leopoldshafen, Germany. E-mail: andreas.barth@fiz-karlsruhe.de

[2] Max Planck Institute for Solid State Research, Heisenbergstraβe 1, D-70569 Stuttgart, Germany. E-mail: w.marx@fkf.mpg.de

[3] Administrative Headquarters of the Max Planck Society, Division for Science and Innovation Studies, Hofgartenstr. 8, 80539 Munich, Germany. E-mail: bornmann@gv.mpg.de

[4] Professorship for Social Psychology and Research on Higher Education, ETH Zurich, Zurich, Switzerland. Email: ruediger.mutz@gess.ethz.ch

**Abstract:**

Subject of our present paper is the analysis of the origins or historical roots of the Higgs boson research from a bibliometric perspective, using a segmented regression analysis in combination with a method named reference publication year spectroscopy (RPYS). Our analysis is based on the references cited in the Higgs boson publications published since 1974. The objective of our analysis consists of identifying specific individual publications in the Higgs boson research context to which the scientific community frequently had referred to. As a consequence, we are interested in seminal works which contributed to a high extent to the discovery of the Higgs boson. Our results show that researchers in the Higgs boson field preferably refer to more recently published papers – particular papers published since the beginning of the sixties. For example, our analysis reveals seven major contributions which appeared within the sixties: Englert & Brout (1964), Higgs (1964, 2 papers), and Guralnik *et al.* (1964) on the Higgs mechanism as well as Glashow (1961), Weinberg (1967), and Salam (1968) on the unification of weak and electromagnetic interaction. Even if the Nobel Prize award highlights the outstanding importance of the work of Peter Higgs and Francois Englert, bibliometrics offer the additional possibility of getting hints to other publications in this research field (especially to historical publications), which are of vital importance from the expert point of view.

## 1. Introduction

On 4 July 2012 the European Organization for Nuclear Research (CERN) announced the observation of a heavy particle around 126 MEV which is supposed to be consistent with the long-sought Higgs boson [1]. The magazine *Science* dedicated several articles to this discovery calling it "the breakthrough of the year" [2]. Even in public newspapers the event was published as the discovery of "God's particle", a name which referred to the title of a book by the Nobel Prize winner Leon Lederman [3]. In March 2013, the detection of the Higgs boson was confirmed by CERN [4]. In October 2013, Peter Higgs and Francois Englert were awarded the Nobel Prize for Physics for their contributions to the standard model of elementary particle physics and the prediction of the boson named after Higgs.

Since the beginning of the 20th century, when Rutherford developed his first atom model, the theory of fundamental particles and their interactions has been a hot topic in physics. One important breakthrough was the development of the unified electromagnetic and weak interaction. Among other ideas, this was based on the concept of broken symmetries and a mechanism for the provision of mass to the otherwise massless vector bosons of the weak interaction, the so-called Higgs mechanism. In 1964, papers on the subject of symmetry breaking and the possibility to create masses for gauge bosons of the weak interaction had been published independently by three research groups (Englert and Brout, [5]; Higgs, [6,7]; Guralnik, Hagen, and Kibble, [8]). According to Close [9], only Higgs "drew attention to the consequential existence of a massive scalar particle, which now bears his name" (p. 141). The other researchers did not mention this boson since "it was obvious" (p. 164. Later, the corresponding field, the mechanism, and the boson were named after Peter Higgs. A few years later, Weinberg [10] and Salam [11] showed that the electromagnetic and weak interactions could be combined into a single theory of the electroweak interaction based on the breakthrough of the Higgs mechanism. At this time they were not aware that Glashow [12] had already developed a theory to solve this problem.

Subject of our present study is the research history of the Higgs boson from a bibliometric perspective. Bibliometric analyses are usually intended to measure the impact of research and they are based on a publication set comprising the publications of a researcher, a research institution, or a journal.

Research performance can be measured by analyzing citation counts of the publications of such sets. Recently, it has been proposed to reverse the perspective of this classic citation analysis from a forward view on the overall citation impact of publications to a backward view on the major contributions to a specific research field [13]. In the latter case the cited references within the publications of a given research field are analyzed in order to determine the importance (the relative "weight") of specific papers, authors, and journals within that field and to quantify their significance.

In a previous analysis, Marx *et al.* [14] have proposed a method to detect the origins or historical roots of research fields by using this backward view of a cited reference analysis. In analogy to classic spectroscopy which shows physical phenomena as peaks in a spectrum, the new method has been named reference publication year spectroscopy (RPYS). RPYS implies to analyze the publication years of the references cited within the body of publications of a specific research field. Major contributions (single frequently referenced publications) appear as prominent peaks in the time series regarding the frequency of the cited references as a function of their publication years (RPYs). As a rule, these few publications are the origins or historical roots of the research field in question.

In this study we identify the origins or historical roots of the Higgs boson research from the perspective of the cited references within the publications of this research field using the segmented regression analysis in the RPYS for the first time. We discuss the results of the RPYS against the backdrop of literature reviews on this field as written, for example, by Close [9] and Bleck-Neuhaus [15].

## 2. Physical Background of Higgs Boson Research

Nature consists of two different types of particles: fermions ("matter") and bosons ("radiation"). Fermions have to occupy different quantum states with the consequence that it is not possible to have two fermions with exactly identical quantum numbers. Examples of fermions are electrons, protons, and neutrons. Bosons, on the other hand, are not subject to such a restriction and they may occupy identical quantum states. The photon ("light") is a boson with zero mass and it is responsible for the electromagnetic interaction. In the second half of the 20[th] century, the standard model evolved as the fundamental model of elementary particle physics. It describes three of the four fundamental forces of nature, *i.e.* electromagnetic, weak, and strong interactions together with the corresponding subatomic particles. So far, gravitation could not be included in this description.

In the standard model the fundamental particles are divided into the three families of fermions (6 leptons and 6 quarks) and the bosons responsible for the electromagnetic (photon), the weak ($W^{\pm}/Z^0$ bosons), and the strong nuclear interactions (gluons). Under local transformations, these physical interactions are invariant, hence the corresponding field theories are gauge-invariant and the carriers of these interactions are called gauge bosons. Since these gauge bosons all have a spin of 1, they are classified as vector bosons, *e.g.* the photon is a massless gauge vector boson. The Higgs boson on the other side has a spin of 0 and is therefore classified as a scalar boson. The $W^{\pm}/Z^0$ bosons are originally massless but they obtain a mass through the (scalar) Higgs boson which interacts with all fundamental particles through the universal Higgs field. The addition of the Higgs boson is required to remove the infinities in the field equations by enabling the formulation of a renormalizable quantum field theory for the electroweak interaction. As a result, the standard model and the Higgs boson are inextricably intertwined. Therefore, it is clear that the detection of the Higgs boson in 2012 was an important milestone for the experimental verification of the standard model ("the breakthrough of the year", [2]).

## 3. Methods

### 3.1 Data

Our analysis is based on the search and retrieval in the database SCISEARCH accessible via STN International. SCISEARCH (the Thomson Reuters Science Citation Index) is a multidisciplinary database covering the publications from core scientific journals together with all the references cited therein. The SCISEARCH database in combination with the search functionalities of STN International enables sophisticated cited reference analyses.

The Higgs boson publications analyzed here were published between 1974 and August 2013 (08-08-2013). SCISEARCH covers no literature prior to 1974. However, for the present study no coverage of former publication years is needed due to the fact that relevant investigation on the Higgs boson which can be searched by corresponding search terms began approximately in the mid1970s. This starting point can be identified with the help of the INSPEC database, which has an excellent coverage of physics literature since 1900 and is also accessible via STN International.

Figure 1 shows the time series of the publications dealing with the standard model (blue), the Higgs boson (red), and related theoretical approaches (string theory, super symmetry, super gravity, and grand unified theory) in summary (green) between 1950 and 2012. The total number of publications in the INSPEC database (grey, divided by a factor of 200) is also included to show the time series of the overall physics literature. It is clearly visible that until the end of the 1970s the number of publications for the standard model, the Higgs boson, and the related theoretical approaches is rather small or even zero. From the late seventies on, the quantity of literature is rising constantly.

In order to reveal the origins or historical roots of the Higgs boson research, we performed a RPYS analysis of the Higgs boson publications covered by SCISEARCH. Even if the publications covered in SCISEARCH are limited to the year 1974, the references cited in the publications are not limited with regard to the reference publication year (RPY). Based on the cited references, we are able to detect those historical publications of outstanding importance for the Higgs boson investigation (and thus figuring as references in publications dealing with the Higgs boson).

The first step of our RPYS was a search for the relevant Higgs boson publications. At the date of searching (08-08-2013), the search for the term "higgs boson(s)" within the titles and abstracts yielded 7,623 papers published since 1974. The second step of the analysis was the extraction of all references cited by the 7,623 Higgs boson publications (altogether 136,403 cited references). Out of the complete set of cited references, the sub-set of 84,678 references belonging to the RPY time period 1900-1990 was analyzed in-depth. The pre-1900 references are much less numerous, much more erroneous, and also much less important here. The past-1990 references refer to more current works which are less important from a historical perspective.

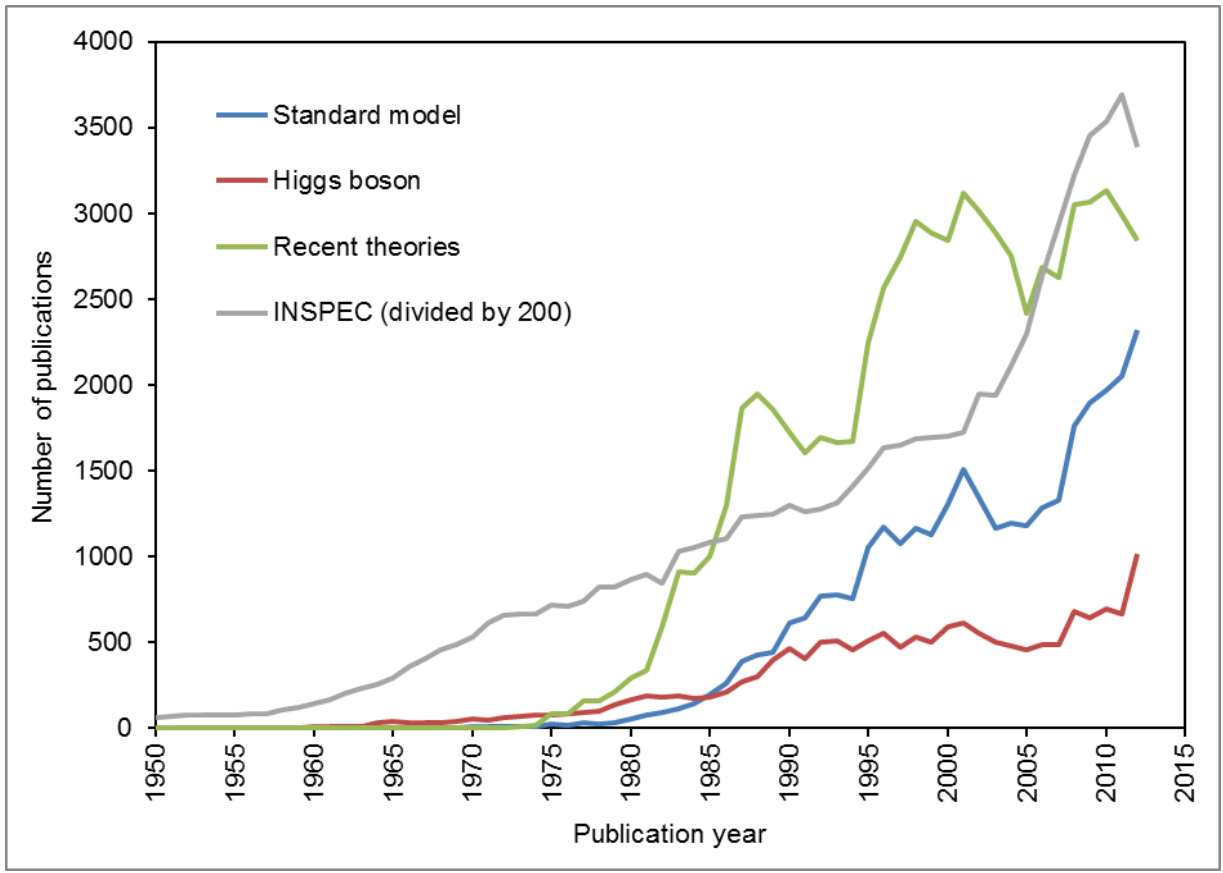

Figure 1: Time series of publications dealing with the standard model (n=32,654), the Higgs boson (n=16,545), related theories (string theory, super symmetry, super gravity, and grand unified theory; n=71,989), and the total number of publications in the database INSPEC (n=14,293,823).

For the RPYS it was necessary to clean the dataset (cited references downloaded from SCISEARCH) with regard to variations of one and the same reference. The references included in citation databases are marginally standardized. In particular, the names of the cited journals may appear written out or may be cited in many possible abbreviations. Furthermore, many references are erroneous (*e.g.* incorrect with regard to the numerical data: volume, starting page, and publication year) [16]. The decisive publication of Abdus Salam (1968) [11] in the Higgs boson research published in the *Proceedings of the Nobel Symposium* is a good example: After collecting all the varying references of this publication, the number of cited references increased from 261 of the most referenced variant to 481 of all variants.

**3.2 Statistical procedure**

In the RPYS of this study, the annual numbers of cited references in the Higgs boson literature were evaluated statistically by regression analysis. To determine different segments of the growth development of cited references within the time series, we used a segmented regression analysis [17-21]. For the analysis, the number of cited references within a year was determined as dependent variable and logarithmized ($\ln(y)$). To eliminate short-term deviations in the time curve of the annual number of cited references, a moving average over five years was calculated. Within the single segments which could be identified by using regression analysis, the references with a high reference volume were fixed. We suppose that these publications are the origins or historical roots of the Higgs boson research field.

For the different segments within the growth development of the annual number of cited references we suppose a simple exponential growth model with $y(t) = y(0) \exp(b_1 t)$, where $b_1$ is the growth constant and $t$ the RPY time. The percent growth rate is given as $(y(t) - y(0)) / y(0)$ is $\exp(b_1) - 1$. We analyze these data with a linear regression model on the variable $\ln(y(t)) = b_0 + b_1 t + \varepsilon$, where $b_0 = \ln(y(0))$ and $\varepsilon$ the residual component. The segmented regression identifies different segments each with individual regression coefficients, respectively, where both the breakpoint (RPY) $a_k$ of the segments as well as the intercept $b_0$ and the growth constant $b_1$ of each segment are estimated. For example, the equation for two segments is ([17], p. 2):

$$\text{IF year} < a_k \text{ THEN } \ln(y) = b_0 + b_1 * \text{year} + \varepsilon \qquad (1)$$

$$\text{ELSE } \ln(y) = b_0 + b_1 * a + b_2 * (\text{year} - a) + \varepsilon$$

$$\varepsilon \sim N(\mathbf{0}, \mathbf{I}\sigma^2),$$

where the residuals $\varepsilon$ are multivariate normally distributed with a zero mean vector and a covariance matrix with identical variances $\sigma^2$ (homoscedasticity) and zero covariances (no autocorrelations of the residuals) overall and across the segments [22, p. 222]. These rather strong assumptions regarding time series data are justified by the fact that given the observed high proportion of explained variance ($R^2 = 0.99$), et vice versa, the low proportion of residual variance, heteroscedasticity or autocorrelation of the residuals are in our case of no importance. Whereas the total variance of ln(y) amounted to 7.19, the overall residual variance amounts to 0.078, and varies across segments from minimal 0.01 (segment 4) to maximal 0.28 (segment 1). The model parameters are estimated by the least squares method (Gauss-Newton) under the restriction that the breakpoints $a_1$, $a_2$, … are ordered. In order to avoid local minima of the estimation procedure, a grid of different starting values for the parameters is used. The regression constant $b_0$ is erased to enhance the fit of the model. The statistical analyses are performed using the SAS procedure PROC NLIN [23].

### 3.3 Results

The results of the segmented regression analysis are shown in table 1. Given the low proportion of residual variance, autocorrelation of the residuals are not of importance. The five segments resulting from the breakpoints in table 1 are as follows: 1900-1915, 1916-1936, 1937-1942, 1943-1949, and 1950-1990. As the growth constants in table 1 show, we have two growth periods within the five segments (1916-1936 and 1950-1990) with a similar growth rate of 13 to 14% interrupted by a break in 1937-1949. This break is caused by the decline of scientific activity around the Second World War.

Table 1: Results of the segmented regression analysis with five segments (1900-1990). The variance explained by the model amounts to $R^2 = 0.99$ or 99%; overall, the model parameters differ statistically significantly from zero according to the F-test: $F(9, 81) = 3,233.4$, $p < 0.05$.

| Parameter | Estimate | Standard error | 95% confidence interval | Growth rate % per RPY |
|---|---|---|---|---|
| Breakpoint | | | | |
| $a_1$ | 1916.3 | 1.08 | [1914.2; 1918.5] | |
| $a_2$ | 1935.9 | 0.55 | [1934.8; 1937.0] | |
| $a_3$ | 1943.3 | 0.47 | [1942.3; 1944.2] | |
| $a_4$ | 1948.8 | 0.99 | [1946.8; 1950.8] | |
| Growth constant | | | | |
| $b_1$ | 0.00 | 0.00 | [0.00; 0.00] | 0.0 |
| $b_2$ | 0.14 | 0.01 | [0.11; 0.16] | 14.6 |
| $b_3$ | -0.28 | 0.04 | [-0.36; -0.19] | -24.2 |
| $b_4$ | 0.42 | 0.09 | [0.25; 0.60] | 52.8 |
| $b_5$ | 0.13 | 0.00 | [0.12; 0.14] | 13.8 |

Figure 2 shows the fitted segmented regression curve. The exponential increase of the annual number of cited references in the five segments identified is the result of two concurrent phenomena. The first phenomenon is the growth of scientific literature: The scientific publications – especially in natural sciences – increased approximately by a factor of hundred throughout the 20th century [16]. In the first half of the 20th century the number of researchers in natural sciences was relatively small. Correspondingly, the number of ensuing publications was low, as reflected in the number of publications for these years [24]. The second phenomenon is named aging (obsolescence, replacement, or oblivion) which means that the interest for scientific papers decreases as time went by [25,26]. Scientists get especially back to publications of recent years and rarely cite publications which had been published many years ago.

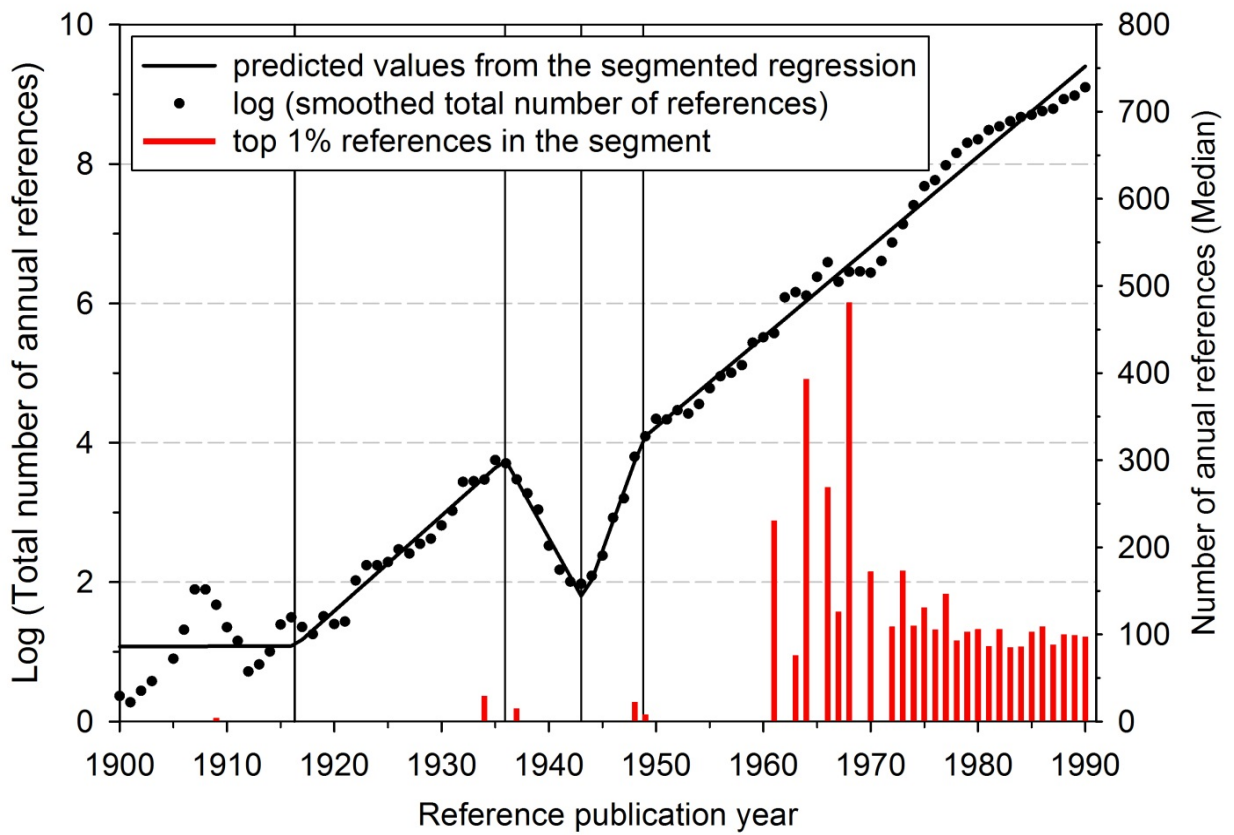

Figure 2: Fitted segmented regression curve. The red bars show the average number of cited references (median) per RPY. The median of those cited references is shown which belong to the top 1% most frequently referenced publications within the five segments identified by regression analysis.

Table 2: The top 1% most frequently referenced publications within the five segments identified by regression analysis (sorted according to reference RPY in ascending order within each segment). For the 5th segment only the 20 most frequently referenced publications from the total of 214 top 1% publications are shown (Appendix A lists all 214 referenced publications).

| No | REF | First author | RPY | Volume | Page | Journal / Book |
|----|-----|--------------|-----|--------|------|----------------|
| **1. Segment** | | | | | | |
| 1 | 4 | NIELSEN N | 1909 | V90 | P123 | NOVA ACTA LEOPOLDIN |
| **2. Segment** | | | | | | |
| 2 | 38 | VON WEIZSACKER C F | 1934 | V88 | P612 | Z PHYS |
| 3 | 21 | WILLIAMS E J | 1934 | V45 | P729 | PHYS REV |
| **3. Segment** | | | | | | |
| 4 | 15 | BLOCH F | 1937 | V52 | P54 | PHYS REV |
| **4. Segment** | | | | | | |
| 5 | 37 | LANDAU L D | 1948 | V60 | P207 | DOKL AKAD NAUK SSSR |
| 6 | 8 | SCHWINGER J | 1948 | V73 | P416 | PHYS REV |
| 7 | 8 | DYSON F J | 1949 | V75 | P1736 | PHYS REV |
| **5. Segment** | | | | | | |
| 8 | 380 | GLASHOW S L | 1961 | V22 | P579 | PHYS |
| 9 | 421 | ENGLERT F | 1964 | V13 | P321 | PHYS REV LETT |
| 10 | 420 | HIGGS P W | 1964 | V12 | P132 | PHYS LETT NUCL |
| 11 | 366 | HIGGS P W | 1964 | V13 | P508 | PHYS REV LETT |
| 12 | 361 | GURALNIK G S | 1964 | V13 | P585 | PHYS REV LETT |
| 13 | 269 | HIGGS P W | 1966 | V145 | P1156 | PHYS REV |
| 14 | 536 | WEINBERG S | 1967 | V19 | P1264 | PHYS REV LETT |
| 15 | 481 | SALAM A | 1968 | | P367 | ELEMENTARY PARTICLE |
| 16 | 270 | COLEMAN S | 1973 | V7 | P1888 | PHYS REV D |
| 17 | 440 | ELLIS J | 1976 | V106 | P292 | NUCL PHYS B |
| 18 | 479 | LEE B W | 1977 | V16 | P1519 | PHYS REV D |
| 19 | 252 | GLASHOW S L | 1977 | V15 | P1958 | PHYS REV D |
| 20 | 294 | GEORGI H M | 1978 | V40 | P692 | PHYS REV LETT |
| 21 | 342 | PASSARINO G | 1979 | V160 | P151 | NUCL PHYS B |
| 22 | 813 | NILLES H P | 1984 | V110 | P1 | PHYS REP |
| 23 | 917 | HABER H E | 1985 | V117 | P75 | PHYS REP |
| 24 | 257 | CHANOWITZ M S | 1985 | V261 | P379 | NUCL PHYS B |
| 25 | 467 | GUNION J F | 1986 | V272 | P1 | NUCL PHYS B |
| 26 | 314 | ELLIS J | 1989 | V39 | P844 | PHYS REV D |
| 27 | 983 | GUNION J F | 1990 | | | HIGGS HUNTERSGUIDE |

Besides the fitted segmented regression curve, figure 2 also shows the distribution of the (non-logarithmic) annual mean number of the references cited by the Higgs boson literature as a bar diagram. The median of those cited references is shown which belong to the top 1% most frequently referenced within the five segments identified by regression analysis. For the total period 1900 to 1990 it becomes apparent that the most frequently referenced publications can be found in the fifth segment (especially between 1960 and 1970).

Whereas the reference counts within the single segments are almost comparable, a comparison on cross segment basis is not possible. Due to the strong increase of the scientific literature, the counts *e.g.* of the references published before the Second World War are not comparable to those references published in the 1970s and 1980s. Another more important reason for the incomparableness is the fact that early publications are no longer explicitly indicated as references, but are taken as known by the reader – this phenomenon is called "obliteration by incorporation" [26]. Thus, table 2 shows the bibliographic data for those publications belonging to the top 1% most referenced publications in its segment being the basis for the calculation of the elevated bar score presented in figure 2. Consequently, the concrete individual publications are listed to which the scientific community particularly often refers within the single segments. These publications are considered to be the origins or historical roots of the Higgs boson research. For the 5th segment only the 20 most frequently referenced publications from the total of 214 top 1% publications are shown in table 2. Appendix A lists all 214 publications.

Within the first four segments, covering the time period 1900-1949, we could identify four publications (no. 2, no. 3, no. 4, and no. 5 in table 2) with more than ten references within the Higgs boson literature: von Weizsäcker (1934, no. 2) [27], Williams (1934, no.3) [28], Bloch (1937, no. 4) [29], and Landau (1948, no. 5) [30] with 38, 21, 15, and 37 references, respectively. These papers deal with "Radiation emitted in collisions with very fast electrons" (von Weizsäcker), with the "Nature of high energy particles" (Williams), with the "Radiation field of the electron" (Bloch), and with "The moment of a 2-photon system" (Landau), analyzing the possibility for the annihilation of slow electrons and positrons. Since all these papers have only a loose connection to the Higgs boson research, it seems that there are scarcely precursor publications in physics or mathematics published before the sixties which can be seen as direct historical roots of the Higgs boson research field. Even the classic papers of quantum physics do not appear extraordinarily frequently as cited references.

According to table 2 (and also figure 2), researchers in the Higgs boson field fall back on publications published since the beginning of the sixties. Among these publications we have those publications by Englert & Brout (1964, no. 9) [5], Higgs (1964, no. 10 and 11) [6,7], and Guralnik *et al.* (1964, no. 12) [8] on the Higgs mechanism as well as the three papers by Glashow (1961, no. 8) [12], Weinberg interaction. According to Close [9, p. 169ff.], the situation in physics was as follows at that time: A quantum field theory of the electromagnetic field had been established and the photon was identified as the corresponding massless gauge vector boson. It was shown that the theory could be renormalized, *i.e.* all infinities disappeared with the right parameter choice. The scientists hoped to create also renormalizable field theories for weak and strong interaction. However, the problem was that Nambu (1961, no. 154) [31] and Goldstone (1961) [32] had shown that a broken symmetry requires the existence of a massless boson (later called Goldstone boson). Such a massless boson could not be found in the experiments and it seemed that nature required only bosons with masses.[1] However, Englert & Brout (1964, no. 9) [5], Higgs (1964, no. 10 and 11) [6,7], and Guralnik *et al.* (1964, no. 12) found a loophole in Goldstone's argument: when the local symmetry is broken, it is possible that massless gauge vector bosons gain mass through the interaction with a new field, later called the Higgs field. This result has been published independently several times within 4 months in the year 1964.

Although Englert & Brout  (1964, no. 9) [5] were the first to describe the new field and the mechanism of symmetry breaking, Higgs [7] was the only one to mention the creation of a massive scalar boson as a result of this process. Hence, his name has been associated with the field, the mechanism, and the boson. The paper by Guralnik *et al.* [8], which was published shortly after the other papers had appeared in 1964 (no. 12), could include references to the other three papers. Together, the four papers contribute with 1,568 out of 1,979 references (79%) to the distinct peak in 1964 in figure 2. The Higgs mechanism was further developed by Higgs (1966, no. 13) [33] and Kibble (1967, no. 78 in Appendix A) [34]. Based on these ideas, Weinberg (1967, no. 14) [10] developed his model of the electro-weak interaction. The theory required two massive charged bosons ($W^+$ and $W^-$) and two neutral bosons – one massive ($Z^0$) plus the massless photon ($\gamma$). Independently of this publication, Glashow (1961, no. 8) [12] and Salam & Ward (1964) [35] came to the same results. Weinberg (1967, no. 14) [10] was able to estimate the masses of the $W^\pm$ and $Z^0$ vector bosons [9, p. 285].

---

[1] This is a possible reason why the paper by Goldstone (1961) is not listed among the most frequently referenced publications in table 2 and Appendix A.

Weinberg (1967, no. 14) [10] also assumed that his model was renormalizable but he did not show this. He referenced Glashow (1961, no. 8) [12] as a predecessor to his own work but not Salam & Ward (1964) [35]. Salam [11] had worked independently of the other authors on a unification of these two fundamental forces (electromagnetic and weak interaction). He presented his work at a conference in Sweden where it was published in the proceedings of the Nobel Symposium [11]. Glashow, Weinberg, and Salam shared the Nobel Prize for Physics in 1979 "for their contributions to the theory of the unified weak and electromagnetic interaction between elementary particles, including, inter alia, the prediction of the weak neutral current" [36]. A prove for the renormalizability of the theory of electroweak interaction was given later by t'Hooft and Veltman (1972, no. 94 in Appendix A) [37] for which they shared another Nobel Prize in 1999 "for elucidating the quantum structure of electroweak interactions in physics" [38].

Since 1976, further frequently referenced publications have been identified by the RPYS (see table 2). Two of these papers (Nilles, 1984, no. 22; Haber, 1985, no. 23) [39,40] are extensive reviews. This indicates that since the mid-eighties the Higgs boson theory has become a major research field which can serve as a basis for a literature summary in form of reviews. With regard to the relatively high reference counts of both reviews, it should be considered that reviews compared to classic articles are referenced above average [41]. The other papers in the table discuss the Higgs boson mass and the possibilities to detect them experimentally as well as the properties of the Higgs boson in a supersymmetric model. Ellis (1976, no. 17) [42] provides a phenomenological profile of the Higgs boson and discusses different possibilities to detect Higgs bosons depending on the possible mass ranges. Lee (1977, no. 18) [43] analyses the role of the Higgs boson mass for the weak interaction at very high energies within the Weinberg-Salam model and Gunion (1986, no. 25) [44] discusses "Higgs Bosons in Supersymmetric Models".

Finally, it should be noted that P.W. Anderson (1963) [45] was the first who discussed the possibility that photons could gain mass by entering a plasma or a superconductor [9, p. 135 ff]. In the theory of superconductivity there is no Goldstone boson present and the photon acts as if it has gained a mass. Anderson [45] concluded that Goldstone's argument might not be valid for all cases. However, the theory of superconductivity is a non-relativistic theory and part of solid state research while particle physics is founded on special relativity. Hence, Anderson's paper did not get enough attendance in the particle physics community (13 refs. in Higgs Boson research).

## 4. Discussion

Subject of our present paper is the analysis of the origins or historical roots of the Higgs boson research from a bibliometric perspective, using a segmented regression analysis in a RPYS. Our analysis is based on the references cited within the Higgs boson publications published since 1974. The objective of our analysis consists of identifying concrete individual publications in the Higgs boson research context to which the scientific community frequently had referred to. As a consequence, we were interested in seminal works which contributed to a high extent to the discovery of the Higgs boson. Even if the Nobel Prize award highlights the outstanding importance of the works of Peter Higgs and Francois Englert, bibliometrics offer the additional possibility of getting hints to other publications in this research field (especially to historical publications), which are of vital importance from the expert point of view.

The segmented regression analysis identified five time segments with two main periods (1916-1936 and 1950-1990) interrupted by a break in 1937-1949 caused by the Second World War. The decisive segment turned out to be the period from 1950 to 1990. We identified four important publications, which appeared prior to 1950 and have been referenced more than ten times: von Weizsäcker (1934), Williams (1934), Bloch (1937), and Landau (1948) [27-30]. Beside these papers there are scarcely other precursor publications in physics or mathematics which have been important for the Higgs boson research community. Researchers in the Higgs boson field preferably refer to more recently published papers – particular papers published since the beginning of the sixties. Our analysis revealed seven major contributions which appeared within the sixties: Englert & Brout (1964), Higgs (1964, 2 papers), and Guralnik *et al.* (1964) [5-8] on the Higgs mechanism as well as Glashow (1961), Weinberg (1967), and Salam (1968) [12,10,11] on the unification of weak and electromagnetic interaction. Since 1976, additional frequently referenced publications have been identified. Two papers (Nilles, 1984; Haber, 1985) [39,40] are extensive reviews. This indicates that in the mid-eighties the Higgs boson theory has become a major research field which can serve as a basis for a literature summary in the form of reviews. Three additional papers discuss the Higgs boson mass and the possibilities to detect them experimentally as well as the properties of the Higgs boson in a supersymmetric model.

As a result of this study, the historical publications which have been cited most frequently by Higgs boson researchers could be identified. However, we cannot act on the assumption that all important publications can be identified by RPYS. Intellectual influences are not always manifest in cited references. Therefore, experts in the field are needed to complete data and information where appropriate. The RPYS method reveals the historical publications potentially relevant for the evolution of a specific research field which could be taken into consideration when its history is reviewed. According to ref. [46], bibliometric methods like RPYS are "no substitute for extensive reading and fine-grained content analysis, if someone is truly interested in the intellectual history of a field" (p. 327).

**Appendix A**

The total of the 214 top 1% most frequently cited publications in the 5th segment sorted according to REF (number of cited references within the Higgs boson literature).

| No | REF | First author | PY | Volume | Page | Journal / Book |
|---|---|---|---|---|---|---|
| 1 | 983 | GUNION J F | 1990 | | | HIGGS HUNTERSGUIDE |
| 2 | 917 | HABER H E | 1985 | V117 | P75 | PHYS REP |
| 3 | 813 | NILLES H P | 1984 | V110 | P1 | PHYS REP |
| 4 | 536 | WEINBERG S | 1967 | V19 | P1264 | PHYS REV LETT |
| 5 | 481 | SALAM A | 1968 | | P367 | ELEMENTARY PARTICLE |
| 6 | 479 | LEE B W | 1977 | V16 | P1519 | PHYS REV D |
| 7 | 467 | GUNION J F | 1986 | V272 | P1 | NUCL PHYS B |
| 8 | 440 | ELLIS J | 1976 | V106 | P292 | NUCL PHYS B |
| 9 | 421 | ENGLERT F | 1964 | V13 | P321 | PHYS REV LETT |
| 10 | 420 | HIGGS P W | 1964 | V12 | P132 | PHYS LETT |
| 11 | 380 | GLASHOW S L | 1961 | V22 | P579 | NUCL PHYS |
| 12 | 366 | HIGGS P W | 1964 | V13 | P508 | PHYS REV LETT |
| 13 | 361 | GURALNIK G S | 1964 | V13 | P585 | PHYS REV LETT |
| 14 | 342 | PASSARINO G | 1979 | V160 | P151 | NUCL PHYS B |
| 15 | 314 | ELLIS J | 1989 | V39 | P844 | PHYS REV D |
| 16 | 294 | GEORGI H M | 1978 | V40 | P692 | PHYS REV LETT |
| 17 | 270 | COLEMAN S | 1973 | V7 | P1888 | PHYS REV D |
| 18 | 269 | HIGGS P W | 1966 | V145 | P1156 | PHYS REV |
| 19 | 257 | CHANOWITZ M S | 1985 | V261 | P379 | NUCL PHYS B |
| 20 | 252 | GLASHOW S L | 1977 | V15 | P1958 | PHYS REV D |
| 21 | 243 | MOHAPATRA R N | 1980 | V44 | P912 | PHYS REV LETT |
| 22 | 239 | PESKIN M E | 1990 | V65 | P964 | PHYS REV LETT |
| 23 | 224 | SHER M | 1989 | V179 | P273 | PHYS REP |
| 24 | 220 | CHAMSEDDINE A H | 1982 | V49 | P970 | PHYS REV LETT |
| 25 | 217 | KOBAYASHI M | 1973 | V49 | P652 | PROG THEOR PHYS |
| 26 | 217 | DERENDINGER J P | 1984 | V237 | P307 | NUCL PHYS B |
| 27 | 205 | ELLIS J | 1984 | V238 | P453 | NUCL PHYS B |
| 28 | 202 | EICHTEN E | 1984 | V56 | P579 | REV MOD PHYS |
| 29 | 201 | DIMOPOULOS S | 1981 | V193 | P150 | NUCL PHYS B |
| 30 | 200 | DREES M | 1989 | V4 | P3635 | INT J MOD PHYS A |
| 31 | 199 | CORNWALL J M | 1974 | V10 | P1145 | PHYS REV D |
| 32 | 197 | HALL L | 1983 | V27 | P2359 | PHYS REV D |
| 33 | 196 | BARBIERI R | 1982 | V119 | P343 | PHYS LETT B |
| 34 | 193 | FRERE J M | 1983 | V222 | P11 | NUCL PHYS B |
| 35 | 192 | INOUE K | 1982 | V68 | P927 | PROG THEOR PHYS |
| 36 | 191 | LEE B W | 1977 | V38 | P883 | PHYS REV LETT |
| 37 | 184 | VELTMAN M | 1977 | V8 | P475 | ACTA PHYS POLON |
| 38 | 182 | WILCZEK F | 1977 | V39 | P1304 | PHYSICAL REVIEW LET |
| 39 | 179 | NILLES H P | 1983 | V120 | P346 | PHYS LETT B |
| 40 | 177 | ALVAREZGAUME L | 1983 | V221 | P495 | NUCL PHYS B |
| 41 | 177 | GUNION J F | 1986 | V278 | P449 | NUCL PHYS B |
| 42 | 172 | GLASHOW S | 1970 | V2 | P1285 | PHYS REV D |
| 43 | 169 | SUSSKIND L | 1979 | V20 | P2619 | PHYS REV D |
| 44 | 169 | SIRLIN A | 1980 | V22 | P971 | PHYS REV D |
| 45 | 168 | PATI J C | 1974 | V10 | P275 | PHYS REV D |

| 46 | 167 | CAHN R N | 1984 | V136 | P196 | PHYS LETT B |
|----|-----|----------|------|------|------|-------------|
| 47 | 165 | BARBIERI R | 1988 | V306 | P63 | NUCL PHYS B |
| 48 | 164 | CABIBBO N | 1979 | V158 | P295 | NUCL PHYS B |
| 49 | 164 | BARDEEN W A | 1990 | V41 | P1647 | PHYS REV D |
| 50 | 161 | BRAATEN E | 1980 | V22 | P715 | PHYS REV D |
| 51 | 156 | FAYET P | 1975 | V90 | P104 | NUCL PHYS B |
| 52 | 152 | BARBIERI R | 1988 | V11 | P1 | RIV NUOVO CIMENTO |
| 53 | 148 | MINKOWSKI P | 1977 | V67 | P421 | PHYS LETT B |
| 54 | 148 | GOLDBERG H | 1983 | V50 | P1419 | PHYS REV LETT |
| 55 | 148 | LAHANAS A B | 1987 | V145 | P1 | PHYS REP |
| 56 | 147 | HABER H E | 1979 | V161 | P493 | NUCL PHYS B |
| 57 | 145 | VELTMAN M | 1977 | V123 | P89 | NUCL PHYS B |
| 58 | 144 | ELLIS R K | 1988 | V297 | P221 | NUCL PHYS B |
| 59 | 142 | KUZMIN V A | 1985 | V155 | P36 | PHYS LETT B |
| 60 | 140 | LINDNER M | 1986 | V31 | P295 | Z PHYS C PARTFIELD |
| 61 | 139 | MOHAPATRA R N | 1981 | V23 | P165 | PHYS REV D |
| 62 | 138 | THOOFT G | 1979 | V153 | P365 | NUCL PHYS B |
| 63 | 138 | SCHECHTER J | 1980 | V22 | P2227 | PHYS REV D |
| 64 | 137 | PECCEI R D | 1977 | V38 | P1440 | PHYS REV LETT |
| 65 | 137 | WEINBERG S | 1979 | V19 | P1277 | PHYS REV D |
| 66 | 137 | BUCHMULLER W | 1986 | V268 | P621 | NUCL PHYS B |
| 67 | 136 | KIM J E | 1984 | V138 | P150 | PHYS LETT B |
| 68 | 136 | KUBLBECK J | 1990 | V60 | P165 | COMPUT PHYS COMMUN |
| 69 | 134 | APPELQUIST T | 1980 | V22 | P200 | PHYS REV D |
| 70 | 134 | WITTEN E | 1981 | V188 | P513 | NUCL PHYS B |
| 71 | 134 | BARGER V | 1990 | V41 | P3421 | PHYS REV D |
| 72 | 132 | APPELQUIST T | 1975 | V11 | P2856 | PHYS REV D |
| 73 | 131 | SENJANOVIC G | 1975 | V12 | P1502 | PHYS REV D |
| 74 | 131 | GINZBURG I F | 1983 | V205 | P47 | NUCL INSTRUMMETHOD |
| 75 | 130 | GINZBURG I F | 1984 | V219 | P5 | NUCL INSTRUMMETH A |
| 76 | 129 | LEE T D | 1973 | V8 | P1226 | PHYS REV D |
| 77 | 128 | SHIFMAN M A | 1979 | V30 | P711 | SOV J NUCL PHYS |
| 78 | 126 | KIBBLE T W B | 1967 | V155 | P1554 | PHYS REV |
| 79 | 126 | GIUDICE G F | 1988 | V206 | P480 | PHYS LETT B |
| 80 | 125 | MOHAPATRA R N | 1975 | V11 | P566 | PHYS REV D |
| 81 | 125 | SAKAI N | 1981 | V11 | P153 | Z PHYS C PARTFIELD |
| 82 | 121 | WEINBERG S | 1976 | V36 | P294 | PHYS REV LETT |
| 83 | 121 | WEINBERG S | 1976 | V37 | P657 | PHYS REV LETT |
| 84 | 121 | DAWSON S | 1985 | V249 | P42 | NUCL PHYS B |
| 85 | 120 | HAGIWARA K | 1987 | V282 | P253 | NUCL PHYS B |
| 86 | 117 | DREES M | 1990 | V240 | P455 | PHYS LETT B |
| 87 | 116 | IBANEZ L | 1982 | V110 | P215 | PHYS LETT B |
| 88 | 113 | DICUS D A | 1973 | V7 | P3111 | PHYS REV D |
| 89 | 113 | SHIFMAN M A | 1978 | V78 | P443 | PHYS LETT B |
| 90 | 113 | KANE G L | 1984 | V148 | P367 | PHYS LETT B |
| 91 | 112 | VAINSHTEIN A I | 1979 | V30 | P711 | SOV J NUCL PHYS |
| 92 | 111 | HEWETT J L | 1989 | V183 | P193 | PHYS REP |
| 93 | 110 | GEORGI H | 1974 | V32 | P438 | PHYS REV LETT |
| 94 | 109 | THOOFT G | 1972 | V44 | P189 | NUCL PHYS B |
| 95 | 109 | BOHM M | 1986 | V34 | P687 | FORTSCHR PHYS |
| 96 | 108 | LONGHITANO A C | 1980 | V22 | P1166 | PHYS REV D |
| 97 | 108 | ANTONIADIS I | 1990 | V246 | P377 | PHYS LETT B |
| 98 | 108 | HOLLIK W F L | 1990 | V38 | P165 | FORTSCHR PHYS |

| 99 | 108 | WEINBERG S | 1990 | V42 | P860 | PHYS REV D |
|---|---|---|---|---|---|---|
| 100 | 107 | LEPAGE G P | 1978 | V27 | P192 | J COMPUT PHYS |
| 101 | 106 | MARCIANO W J | 1980 | V22 | P2695 | PHYS REV D |
| 102 | 106 | INOUE K | 1982 | V67 | P1889 | PROG THEOR PHYS |
| 103 | 105 | SIEGEL W | 1979 | V84 | P193 | PHYS LETT B |
| 104 | 104 | IOFFE B L | 1978 | V9 | P50 | SOV J PART NUCL |
| 105 | 103 | DUGAN M | 1985 | V255 | P413 | NUCL PHYS B |
| 106 | 102 | FUKUGITA M | 1986 | V174 | P45 | PHYS LETT B |
| 107 | 102 | CHENG T P | 1987 | V35 | P3484 | PHYS REV D |
| 108 | 101 | GELLMANN M | 1979 | | P315 | SUPERGRAVITY |
| 109 | 101 | LONGHITANO A | 1981 | V188 | P118 | NUCL PHYS B |
| 110 | 101 | MIRANSKY V A | 1989 | V221 | P177 | PHYS LETT B |
| 111 | 100 | GUNION J F | 1988 | V299 | P231 | NUCL PHYS B |
| 112 | 99 | DICUS D A | 1989 | V39 | P751 | PHYS REV D |
| 113 | 99 | MIRANSKY V A | 1989 | V4 | P1043 | MOD PHYS LETTA |
| 114 | 99 | WEINBERG S | 1989 | V63 | P2333 | PHYS REV LETT |
| 115 | 98 | FLORES R A | 1983 | V148 | P95 | ANN PHYS-NEWYORK |
| 116 | 98 | BERGER M S | 1990 | V41 | P225 | PHYS REV D |
| 117 | 97 | GUNION J F | 1990 | V42 | P1673 | PHYS REV D |
| 118 | 96 | FAYET P | 1977 | V69 | P489 | PHYS LETT B |
| 119 | 96 | GLASHOW S L | 1978 | V18 | P1724 | PHYS REV D |
| 120 | 96 | ALVAREZGAUME L | 1982 | V207 | P96 | NUCL PHYS B |
| 121 | 95 | ALTARELLI G | 1977 | V126 | P298 | NUCL PHYS B |
| 122 | 94 | IBANEZ L E | 1984 | V233 | P511 | NUCL PHYS B |
| 123 | 94 | GUNION J F | 1987 | V294 | P621 | NUCL PHYS B |
| 124 | 93 | YANAGIDA T | 1979 | | P95 | P WORKSH UN THEOR B |
| 125 | 93 | CHENG T P | 1980 | V22 | P2860 | PHYS REV D |
| 126 | 93 | BARR S M | 1990 | V65 | P21 | PHYS REV LETT |
| 127 | 92 | KUNSZT Z | 1984 | V247 | P339 | NUCL PHYS B |
| 128 | 92 | GORISHNY S G | 1990 | V5 | P2703 | MOD PHYS LETTA |
| 129 | 90 | WEINBERG S | 1976 | V13 | P974 | PHYS REV D |
| 130 | 90 | CHANOWITZ M S | 1978 | V78 | P285 | PHYS LETT B |
| 131 | 90 | CHANOWITZ M S | 1979 | V153 | P402 | NUCL PHYS B |
| 132 | 90 | VELTMAN M | 1981 | V12 | P437 | ACTA PHYS POLB |
| 133 | 90 | GRAY N | 1990 | V48 | P673 | Z PHYS C |
| 134 | 89 | PECCEI R D | 1977 | V16 | P1791 | PHYS REV D |
| 135 | 89 | JONES D R T | 1979 | V84 | P440 | PHYS LETT B |
| 136 | 89 | GIRARDELLO L | 1982 | V194 | P65 | NUCL PHYS B |
| 137 | 88 | FARHI E | 1981 | V74 | P277 | PHYS REP |
| 138 | 88 | GUNION J F | 1989 | V39 | P2701 | PHYS REV D |
| 139 | 86 | WEINBERG S | 1979 | V96 | P327 | PHYSICA A |
| 140 | 86 | MANOHAR A | 1984 | V234 | P189 | NUCL PHYS B |
| 141 | 85 | VAYONAKIS C E | 1976 | V17 | P383 | LETT NUOVO CIMENTO |
| 142 | 85 | FAYET P | 1977 | V32 | P249 | PHYS REP C |
| 143 | 85 | DIMOPOULOS S | 1981 | V24 | P1681 | PHYS REV D |
| 144 | 85 | AOKI K | 1982 | V73 | P1 | PROG THEOR PHYS SUP |
| 145 | 85 | DASHEN R | 1983 | V50 | P1897 | PHYS REV LETT |
| 146 | 85 | INOUE K | 1984 | V71 | P413 | PROG THEOR PHYS |
| 147 | 85 | KAPLAN D B | 1984 | V136 | P183 | PHYS LETT B |
| 148 | 84 | NASON P | 1988 | V303 | P607 | NUCL PHYS B |
| 149 | 83 | COLLINS J C | 1985 | V250 | P199 | NUCL PHYS B |
| 150 | 82 | LINDE A D | 1976 | V23 | P64 | JETP LETT |
| 151 | 82 | HILL C T | 1981 | V24 | P691 | PHYS REV D |

| 152 | 82 | ALTARELLI G | 1987 | V287 | P205 | NUCL PHYS B |
| 153 | 82 | HOLDOM B | 1990 | V247 | P88 | PHYS LETT B |
| 154 | 81 | NAMBU Y | 1961 | V122 | P345 | PHYS REV |
| 155 | 81 | WESS J | 1974 | V70 | P39 | NUCL PHYS B |
| 156 | 81 | GELMINI G B | 1981 | V99 | P411 | PHYS LETT B |
| 157 | 81 | GEORGI H | 1985 | V262 | P463 | NUCL PHYS B |
| 158 | 80 | MOHAPATRA R N | 1975 | V11 | P2558 | PHYS REV D |
| 159 | 80 | SHAPOSHNIKOV M E | 1987 | V287 | P757 | NUCL PHYS B |
| 160 | 79 | CAPPER D M | 1980 | V167 | P479 | NUCL PHYS B |
| 161 | 79 | LAZARIDES G | 1981 | V181 | P287 | NUCL PHYS B |
| 162 | 78 | DESHPANDE N G | 1978 | V18 | P2574 | PHYS REV D |
| 163 | 78 | DIMOPOULOS S | 1979 | V155 | P237 | NUCL PHYS B |
| 164 | 78 | BIGI I | 1986 | V181 | P157 | PHYS LETT B |
| 165 | 78 | BARNETT R M | 1988 | V306 | P697 | NUCL PHYS B |
| 166 | 77 | SAKHAROV A D | 1967 | V5 | P24 | JETP LETT |
| 167 | 77 | BARDEEN W A | 1978 | V18 | P3998 | PHYS REV D |
| 168 | 77 | ELLIS J | 1983 | V121 | P123 | PHYS LETT B |
| 169 | 77 | AMALDI U | 1987 | V36 | P1385 | PHYS REV D |
| 170 | 76 | CABIBBO N | 1963 | V10 | P531 | PHYS REV LETT |
| 171 | 76 | DOLAN L | 1974 | V9 | P3320 | PHYS REV D |
| 172 | 75 | EICHTEN E | 1980 | V90 | P125 | PHYS LETT B |
| 173 | 73 | KAPLAN D B | 1984 | V136 | P187 | PHYS LETT B |
| 174 | 73 | GUNION J F | 1989 | V40 | P1546 | PHYS REV D |
| 175 | 72 | PENDLETON B | 1981 | V98 | P291 | PHYS LETT B |
| 176 | 72 | ELLIS J | 1983 | V128 | P248 | PHYS LETT B |
| 177 | 72 | KEUNG W Y | 1984 | V30 | P248 | PHYS REV D |
| 178 | 72 | BEENAKKER W | 1989 | V40 | P54 | PHYS REV D |
| 179 | 71 | FARRAR G R | 1978 | V76 | P575 | PHYS LETT B |
| 180 | 71 | FLEISCHER J | 1981 | V23 | P2001 | PHYS REV D |
| 181 | 71 | POLCHINSKI J | 1983 | V125 | P393 | PHYS LETT B |
| 182 | 71 | SJOSTRAND T | 1987 | V43 | P367 | COMPUT PHYS COMMUN |
| 183 | 71 | LINDNER M | 1989 | V228 | P139 | PHYS LETT B |
| 184 | 71 | BAUR U | 1990 | V339 | P38 | NUCL PHYS B |
| 185 | 70 | ELLIS J | 1979 | V83 | P339 | PHYS LETT B |
| 186 | 70 | COLLINS J C | 1981 | V193 | P381 | NUCL PHYS B |
| 187 | 70 | ELLIS J | 1982 | V114 | P231 | PHYS LETT B |
| 188 | 70 | GASSER J | 1984 | V158 | P142 | ANN PHYS-NEWYORK |
| 189 | 70 | VANOLDENBORGH G | 1990 | V46 | P425 | Z PHYS C PARTFIELD |
| 190 | 69 | KOUNNAS C | 1984 | V236 | P438 | NUCL PHYS B |
| 191 | 69 | IBANEZ L E | 1985 | V256 | P218 | NUCL PHYS B |
| 192 | 69 | KLEISS R | 1985 | V262 | P235 | NUCL PHYS B |
| 193 | 69 | KUTI J | 1988 | V61 | P678 | PHYS REV LETT |
| 194 | 69 | TELNOV V I | 1990 | V294 | P72 | NUCL INSTRUMMETH A |
| 195 | 68 | WILCZEK F | 1978 | V40 | P279 | PHYS REV LETT |
| 196 | 68 | MAGG M | 1980 | V94 | P61 | PHYS LETT B |
| 197 | 68 | BUCHMULLER W | 1983 | V121 | P321 | PHYS LETT B |
| 198 | 68 | MACHACEK M E | 1983 | V222 | P83 | NUCL PHYS B |
| 199 | 68 | GLOVER E W N | 1988 | V309 | P282 | NUCL PHYS B |
| 200 | 68 | BAWA A C | 1990 | V47 | P75 | Z PHYS C PARTFIELD |
| 201 | 67 | FROGGATT C D | 1979 | V147 | P277 | NUCL PHYS B |
| 202 | 67 | VAINSHTEIN A I | 1980 | V23 | P429 | SOV PHYS USP |
| 203 | 66 | SAKAI N | 1980 | V22 | P2220 | PHYS REV D |
| 204 | 66 | LI S P | 1984 | V140 | P339 | PHYS LETT B |

| 205 | 66 | GOUNARIS G J | 1986 | V34 | P3257 | PHYS REV D |
| 206 | 66 | VOLOSHIN M B | 1986 | V44 | P478 | SOV J NUCL PHYS+ |
| 207 | 66 | GAMBERINI G | 1990 | V331 | P331 | NUCL PHYS B |
| 208 | 65 | ELLIS J | 1983 | V125 | P275 | PHYS LETT B |
| 209 | 65 | MACHACEK M E | 1984 | V236 | P221 | NUCL PHYS B |
| 210 | 65 | BAGGER J | 1990 | V41 | P264 | PHYS REV D |
| 211 | 64 | CAHN R N | 1979 | V82 | P113 | PHYS LETT B |
| 212 | 64 | INOUE K | 1983 | V70 | P330 | PROG THEOR PHYS |
| 213 | 64 | HALL L J | 1984 | V231 | P419 | NUCL PHYS B |
| 214 | 64 | KLINKHAMER F R | 1984 | V30 | P2212 | PHYS REV D |